\documentclass{pasa}%

\title[GLACiAR]{GLACiAR, an open-source python tool for simulations of source recovery and completeness in galaxy surveys.}
\author[Carrasco, D. et al.]{D. Carrasco$^1$\thanks{dcarrasco@student.unimelb.edu.au}, M. Trenti$^{1,2}$, S. Mutch$^{1,2}$, \and P.~A. Oesch$^{3,}$\\
\affil{$^1$School of Physics, The University of Melbourne, Parkville, VIC 3010, Australia.}%
\affil{$^2$ ARC Centre of Excellence for All-Sky Astrophysics in 3 Dimensions (ASTRO-3D), Australia.}%
\affil{$^3$Observatoire de Gen\'eve, 51 Ch. des Maillettes, CH-1290 Versoix, Switzerland.}%
}%
\jid{PASA}
\doi{10.1017/pas.\the\year.xxx}
\jyear{\the\year}

\usepackage[authoryear]{natbib}
\bibpunct{(}{)}{;}{a}{}{,}
\usepackage{aas_macros}
\usepackage{hyperref} 
\hypersetup{colorlinks,citecolor=blue,linkcolor=blue,urlcolor=blue}

\begin{document}%
\begin{abstract}
The luminosity function is a fundamental observable for characterizing
how galaxies form and evolve throughout the cosmic history. One key
ingredient to derive this measurement from the number counts in a
survey is the characterization of the completeness and redshift
selection functions for the observations. In this paper we present \texttt{GLACiAR}, an
open \texttt{python} tool available on \texttt{GitHub} to estimate the
completeness and selection functions in galaxy surveys. The code is
tailored for multiband imaging surveys aimed at searching for
high-redshift galaxies through the Lyman Break technique, but it can
be applied broadly. The code generates artificial galaxies that
follow S\'ersic profiles with different indexes and with
customizable size, redshift and spectral energy distribution
properties, adds them to input images, and measures the recovery
rate. To illustrate this new software tool, we apply it to quantify
the completeness and redshift selection functions for J-dropouts
sources (redshift $z\sim 10$ galaxies) in the Hubble Space Telescope Brightest of
Reionizing Galaxies Survey (BoRG). Our comparison with a previous completeness analysis on the same dataset shows overall agreement, but also highlights how different modelling assumptions for artificial sources can impact completeness estimates.
\end{abstract}
\maketitle%
\section{INTRODUCTION}

The search for high redshift objects has rapidly developed in the last decades as astronomers attempt to understand the evolution of galaxies throughout the history of the Universe, with the current frontier being at redshift $z\sim 10$, or $\sim13.4$ Gyr lookback time {\citep{oesch2016, zitrin2014, coe2013}}. Since the large majority of these distant sources are very faint {($m_{AB}\sim26$ for a typical $L_*$ galaxy at $z\sim6$)} deep images of the sky are needed. The \textit{Hubble Space Telescope} has carried out a number of surveys that had the detection of high-redshift galaxies as a key science motivation, starting from the pioneering \textit{Hubble} Deep Field survey \citep[HDF; ][]{hdf}, and then continuing to improve depth and area covered thanks to technological progress offering newer instrumentation, with the \textit{Hubble} Ultra Deep Field survey \citep[HUDF; ][]{hudf}, the Great Observatories Origins Deep Survey \citep[GOODS; ][]{goods}, {the Cosmic Assembly Near-infrared Deep Extragalactic Legacy Survey \citep[CANDELS; ][]{candels}, the \textit{HST} Frontier Fields \citep{hstff}, and the Brightest of Reionizing Galaxies Survey \citep[BoRG;][]{trenti2011}, among others.}   

The most common techniques used to identify high redshift galaxies from broadband imaging are the Lyman-break method \citep{steidel96}, which has been widely applied to the highest redshift ($z\gtrsim4$) samples {\citep[e.g.][]{bouwens2015}}, and other photometric redshift selection methods \citep[e.g.][]{coe2006}. Due to the ubiquitous presence of hydrogen, which has a large ionization cross section, photons with $\lambda < 912$\AA{} are heavily absorbed by neutral hydrogen in its ground state, and only have a low probability of escaping from a galaxy without being absorbed. Hydrogen in the intergalactic medium also contributes to the Lyman-break, effectively absorbing a large fraction of photons emitted by a high-redshift source at $\lambda < 1216$\AA{} for sources at $z\gtrsim 4$. Although generally highly effective, the Lyman-break method has some limitations as it may preferentially select only certain subsets of the galaxy population at high-$z$, such as relatively unobscured, actively star-forming galaxies (e.g., see \citealt{stanway2008}). Recent examples of the application of this technique include \citet{calvi2016}, \citet{bouwens2016}, and \citet{hathi2010}. The Lyman break selection is a special case of multi-color photometric selection, which is most effective when a spectral break is present in the sources targeted by the survey. However, spectra of galaxies can have other characteristic features in addition to the Lyman-break, which can be observed in different wavelengths and can improve the candidates selection. For instance, infrared data can be used to detect the Balmer break in $z\gtrsim5$ galaxies \citep{mobasher2005}, and photometric redshift accuracy and reliability improves when there is a large number of bands available. 

Arguably, one of the most fundamental observables from high-redshift surveys is the measurement of the galaxy luminosity function (LF). Generally, studies of the LF at cosmological distances are carried out with galaxy candidates from photometric catalogs (either using photometric redshift estimations or a dropout technique) as spectroscopic samples are significantly more challenging to construct and thus limited in numbers. Even after accounting for the most recent advancements in the field, that yielded catalogs of photometric sources at $z\gtrsim 4$ including more than $10,000$ sources \citep{bouwens2015}, the LF shape is still debated, and the topic is a very active research area \citep[e.g.][]{ishigaki2017, bouwens2015, atek2015, bowler2014, bradley2012}. To go from counting galaxies to the construction of the LF, it is imperative to understand completeness and efficiency, i.e., what fraction of all the existing galaxies with a given spectral energy distribution, morphological properties and redshift range is identified in an observed sample. Accordingly, a machinery able to estimate the recovery fraction is critically needed for robust LF estimations. Yet, despite the large number of high-redshift galaxy surveys carried out in the last 20 years since the original Hubble Deep Field \citep{williams96}, there is not a unified publicly available tool to estimate their completeness and source recovery. Such a software tool is not only important for the estimation of volume and luminosity functions, but also to investigate the properties of the galaxies a survey fails to detect, and reasons for missing them. 

The classic approach to completeness estimates is to insert simulated galaxies in the observed images and quantify the recovery efficiency. There are two main methods typically used to create these simulated sources. One is based on starting from images of galaxies acquired in similar observations (for example at lower redshift), that are modified/re-scaled to fit the desired properties of the sample to simulate. The other one is the creation of artificial light profiles from theoretical models of the expected surface brightness profiles. Examples of LF studies utilizing the former approach are \citet{bershady98}, \citet{imai2007}, and \citet{hornillos2009}. The latter approach is applied in \citet{bowler2015}, \citet{oesch2014}, \citet{jian2011}, among others.
This is also the approach taken in this paper, primarily because of its flexibility in the definition of shape, size and the spectral energy distribution of the artificial sources, which make it well suited for a broader range of applications.

This paper presents a \texttt{python} based tool to estimate the completeness of galaxy surveys, the \emph{GaLAxy survey Completeness AlgoRithm} (\texttt{GLACiAR} hereafter). The software produces a photometric output catalog of the simulated sources as main output, and associated higher level products to easily quantify source completeness and recovery. In particular, two main analyses are automatically performed: the first is the calculation of the fraction of sources recovered as a function of magnitude in the detection band (i.e., the survey completeness); and the second one is a more comprehensive characterization of the recovery efficiency taking into account all survey bands allowing the user to implement multi-color selection criteria to identify high-redshift galaxies (i.e., the survey source selection efficiency as a function of both input magnitude and redshift).

The current version of the software is limited to handle blank (non-lensed) fields, but the code structure has been designed with the idea of introducing, in a future release, the capability to load a user-defined lensing magnification map and add artificial objects in the source plane. This would allow natural application of the code to quantify completeness for lensing surveys, which is a powerful complementary method to find high-redshift galaxies as we can observe intrinsically faint galaxies that are magnified by foreground objects. Surveys such as the Cluster Lensing And Supernova survey with \textit{Hubble} \citep[CLASH;][]{clash} and the Herschel Lensing Survey \citep[HLS;][]{hls} are some examples. 

This paper is organized as follows: Section~\ref{overview} discusses the principles of the code, with our specific algorithmic implementation presented  in Section~\ref{implementation}. Section~\ref{results} illustrates the application of the code to part of the BoRG survey and compares the results obtained to previous determinations of the survey completeness and selection functions. Finally, we summarize and conclude in Section~\ref{discussion}.

\section{GENERAL OVERVIEW}\label{overview}

\begin{figure*}
    \centering
    \includegraphics[scale=0.3]{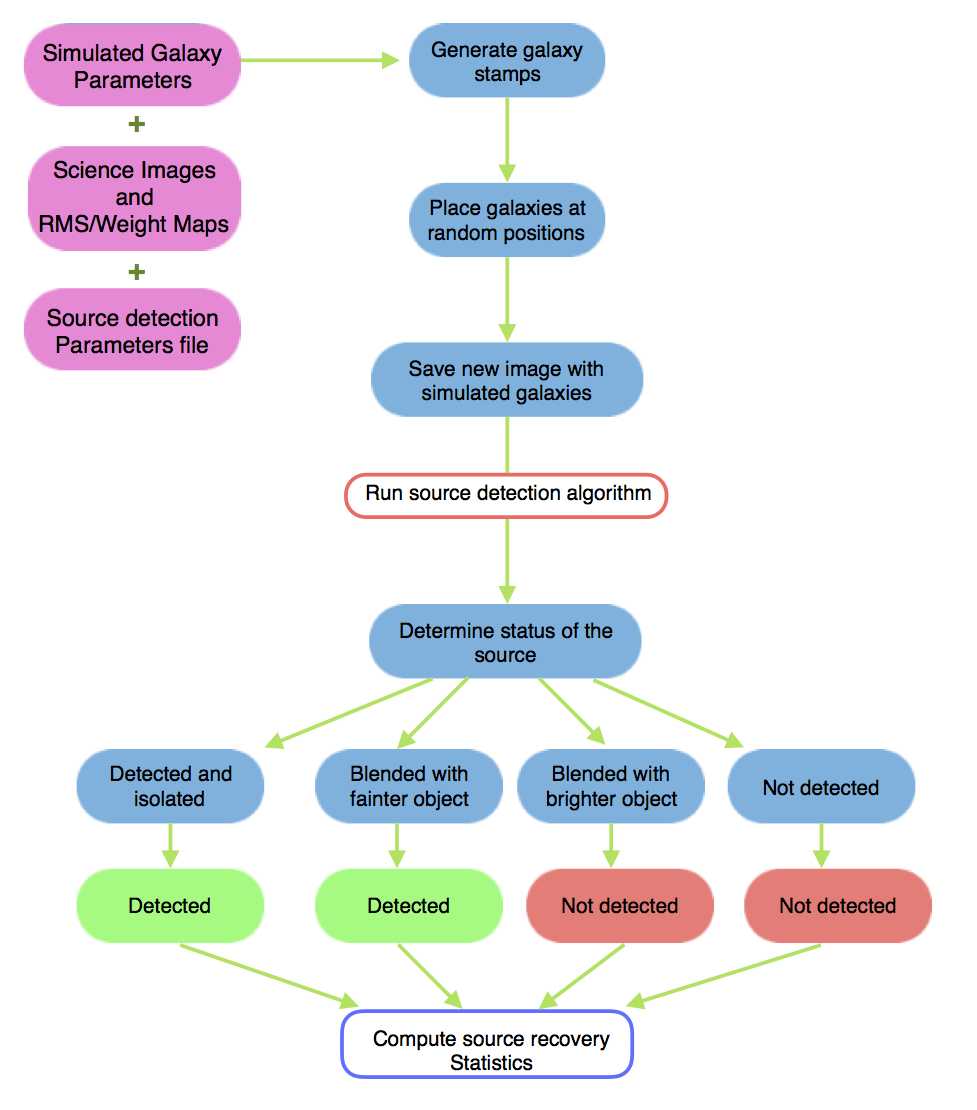}
    \caption{Logic diagram of \texttt{GLACiAR}'s code structure. User-defined parameters and a science image (with its associated RMS map) are taken as input, with the code then generating simulated galaxy stamps, which are added to the science image at random positions, sampled from a uniform distribution. A detection algorithm is run on these images, and its output is used to determine statistics on source recovery.}
    \label{fig:diagram}
\end{figure*}

\texttt{GLACiAR} is structured modularly for maximum efficiency and flexibility. First, it creates artificial galaxies and adds them to the observed science images. Then, a module to identify sources is called, which builds catalogs with photometric information of the detected objects. The output catalogs from the original science images are compared with the ones from the new frames in order to identify the artificial sources recovery and multi-band photometric information. Finally, another module is available to automatically calculate their recovered fraction as a function of input magnitude and simulated redshift. Figure~\ref{fig:diagram} provides a high-level summary of the algorithm.

To identify sources, we limit ourselves to the use of \texttt{SExtractor} \citep{sextractor} for the current release, but we expect to expand the functionality of \texttt{GLACiAR} to allow the use of \texttt{photutils} \citep{photutils} in future versions.

A set of galaxy stamps are generated with sources that follow a S\'ersic luminosity distribution \citep{sersic68} with parameters defined by the user. These artificial galaxies are placed at random positions on the images of the survey. In order to run the code, a parameters file (described in Section~\ref{input_parameters}) must be completed by the user to define the features of the simulated galaxies, such as magnitude, size, redshift, among others. Along with this, \texttt{GLACiAR} requires other files: the science images, a list with the names of the fields (for one or more than one), the \texttt{SExtractor} parameters, frames with noise intensity maps (RMS or weight maps, depending on which ones are used to run the source identification), and the point spread functions (PSFs) in the filter(s) used to acquire the image(s). These inputs are described in more detail in Section~\ref{files_needed}.

\subsection{S\'ersic profiles for artificial galaxies}

For the characterization of the artificial galaxy's surface brightness, we use the S\'ersic luminosity profile \citep{sersic68} which has been widely shown to be a good fit for different types of galaxies given its flexibility \citep[e.g.:][]{peng2002, graham2005, haussler2013}. This profile is defined as:
\begin{equation}
I(R) = I_{e}exp\left \{-b_{n}\bigg[\bigg(\frac{R}{R_e}\bigg)^{\frac{1}{n}}-1\bigg]\right \},
\label{eq1}
\end{equation}
with $I_{e}$ being the intensity at the radius that encloses half of the total light, $R_{e}$; $n$ is the S\'ersic index, which describes the shape of the profile; and $b_{n}$ is a constant defined in terms of this index, which follows from our choice to normalize the profile with $I_e$.

To obtain the luminosity of a galaxy within a certain radius, we follow the approach by \citet{graham2005} integrating equation (\ref{eq1}) over a projected area $A=\pi R^{2}$, ending up with:
\begin{equation}
L(<R) = I_{e}R_{e}^{2}2\pi n\frac{e^{b_{n}}}{(b_{n})^{2n}}\gamma(2n,x)\\
\label{eq2}
\end{equation}
where $\gamma(2n,x)$ is the incomplete gamma function, and
\begin{equation}
x = b_{n} \bigg(\frac{R}{R_{e}}\bigg)^{\frac{1}{n}}.
\end{equation}
To calculate $b_{n}$ we follow \citet{ciotti91}, and taking the total luminosity we obtain:
\begin{equation}
\Gamma(2n) = 2\gamma(2n,b_{n})
\end{equation}
where $\Gamma$ is the complete gamma function. From here, the value of $b_{n}$ can be obtained.

\subsection{Artificial galaxy data}\label{artificialgalaxy}

We create the stamp of an artificial galaxy according to a set of input (user-specified) parameters, which describe the free parameters of the S\'ersic profile described in equation~\ref{eq2}. $n$ is the S\'ersic index and it can be defined arbitrarily in \texttt{GLACiAR}. For the effective radius $R_{e}$, the input is the proper size in kiloparsecs at a redshift $z=6$, which is converted into arcseconds and scaled by the redshift with $(1+z)^{-1}$. The default value is $R_{e}=1.075$ kpc, chosen according to previous completeness simulations for the BoRG survey \citep{bradley2012, bernard}. This is converted into arcseconds by using the scale of the images. The intensity $I_{e}$ is calculated from equation $\ref{eq2}$ considering $L(<R)$ as the total flux, which depends on the magnitude assigned to the object. Each magnitude can be converted into flux using
\begin{equation}
f_{b} = 10^{\frac{(zp_{b}-m_{b})}{2.5}},
\end{equation}
with $f_{b}$, $zp_{b}$, and $m_{b}$ being the flux, zeropoint, and magnitude of a $``b"$ band, respectively. The user specifies the value for the magnitude in the detection band (which is also chosen by the user). The flux in the other bands is calculated according to the redshift of the simulated galaxy and its spectrum. To calculate the flux in each filter and for each object, we assume a power-law spectrum with a Lyman break as a function of the wavelength $\lambda$:
\begin{equation}
F(\lambda) = \left\{
        \begin{array}{ll}
            0 & \quad \lambda \leq 0 \\
            a\lambda^{\beta} & \quad 1216\leq \lambda
        \end{array}
    \right.,
\label{eq_spectrum}
\end{equation}
where $a$ is the normalization, and $\beta$ is the slope of the flux. In our code, the value of $\beta$ follows a Gaussian distribution, where the mean and standard distribution can be chosen by the user. For the default case, we adopt a mean of $-2.2$ and a standard deviation of $0.4$, which is suitable for high redshift galaxies \citep{bouwens2015}. 

\begin{figure*}
    \centering
    \includegraphics[scale=0.6]{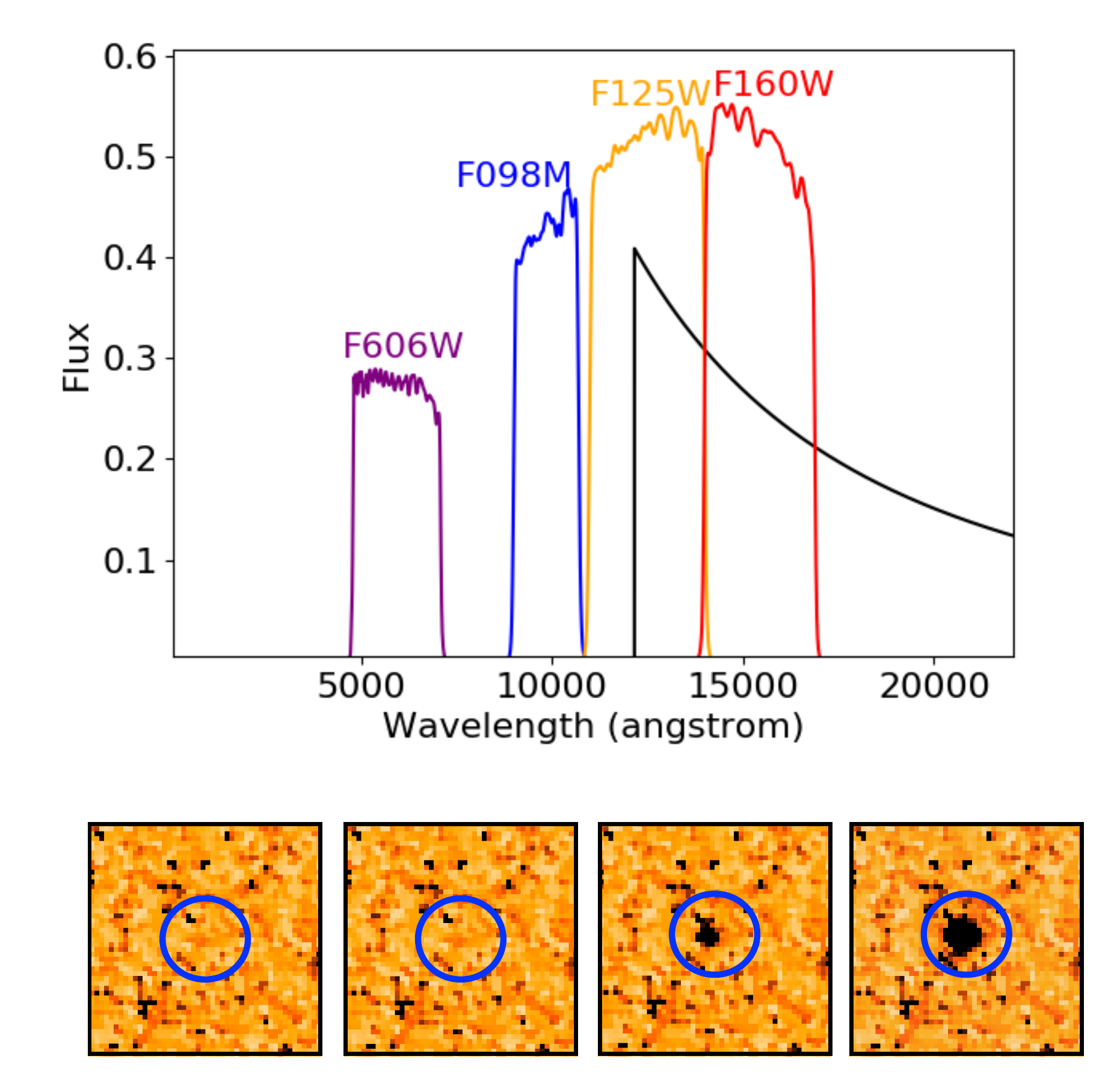}
    \caption{\textit{Top:} Spectrum of a simulated galaxy at $z=10$ and with $\beta=-2.0$ produced by \texttt{GLACiAR} in arbitrary units of flux as a function of wavelength, with four \textit{HST} filter transmission curves superimposed (F098M, F125W, F160W, and F606W). \textit{Bottom:} Source from above inserted into the F606W, F098M, F125W, and F160W science images (from left to right) from field BoRG-0835+2456 assuming a $n=4$ surface brightness profile and $m_{AB}=24.0$ with no inclination and circular shape. The stamps have a size of 3.6''$\times$3.6''.}
    \label{spectrum}
\end{figure*}

In the top panel of Figure~\ref{spectrum} we show the spectrum of a simulated galaxy with $\beta = -2.0$ at $z=10.0$ with the filters F098M, F125W, F160W, and F606W from \textit{HST} used in the BoRG survey (described in section~\ref{borg}). The bottom panel shows that same source added to the science images in those filters. It can be seen that there is no image of the artificial galaxy in the F098M and F125W bands, as no flux is expected at these wavelengths, while the artificial source is present in F125W and F160W bands with different intensities, since the Lyman-break falls in the F125W filter. 

\begin{figure*}
    \centering
    \includegraphics[scale=0.55]{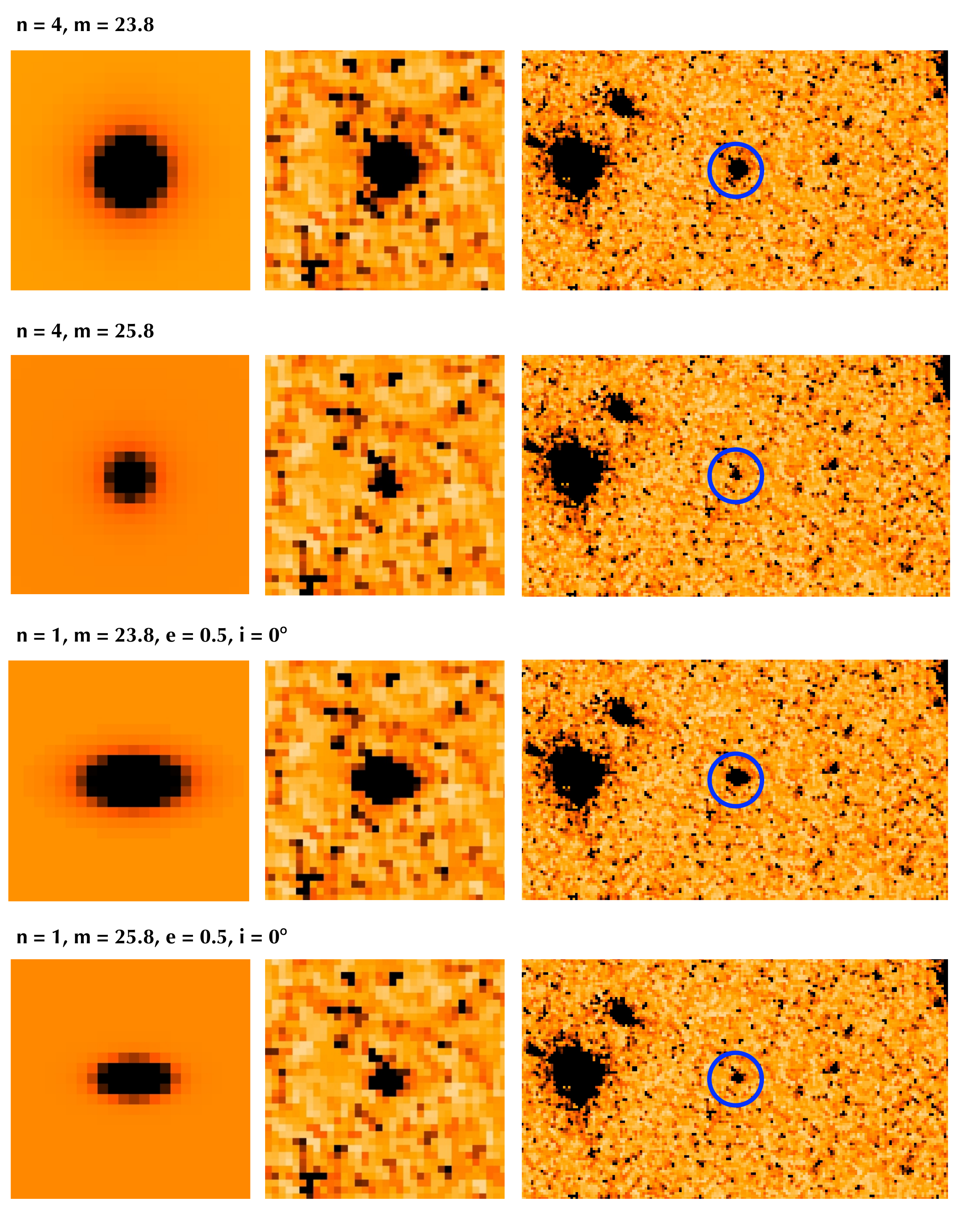}
    \caption{Example of different types of galaxies produced by \texttt{GLACiAR}. The left panels show a zoom of the galaxies placed on a constant background (box size 35$\times$35 pixels), while the middle and right panels show them inserted in a typical science image (F160W for the field BoRG-0835+2456) with box sizes (2.8''$\times$2.8'' and 5.0''$\times$2.8'' respectively). From top to bottom we see an artificial galaxy with a S\'ersic index of 4, and total input magnitude $m_{AB}=23.8$; an artificial galaxy with S\'ersic index of 4, and magnitude $m_{AB}=25.8$; an artificial galaxy with S\'ersic index of 1, magnitude $m_{AB}=23.8$, eccentricity of 0.5, and inclination angle of 05$^{\circ}$; and an artificial galaxy with S\'ersic index of 1, magnitude $m_{AB}=25.8$, eccentricity of 0.5, and inclination angle of 0$^{\circ}$. The first two ones have a circular shape, while the latter two are elliptical.}
    \label{galaxy_stamps}
\end{figure*}

The user can choose different S\'ersic indexes for the simulated galaxies as well as the fraction of each type. The default values are $50\%$ of the sources with $n=1$, and $50\%$ with $n=4$.
In terms of morphology, the galaxies can have different inclinations and eccentricities. The inclinations can vary from $0^{\circ}$ to $90^{\circ}$, and the user can specify the sampling sequence in the angular coordinate space. For example, if 10 values are chosen, the sampling spacing will be $9^{\circ}$. The same principle applies to eccentricities, whose values vary from 0 (circle) to almost 1 (highly elliptical). Furthermore, we allow for a special case: a S\'ersic index of $n=4$. This profile \citep{devaucouleurs} is commonly associated with elliptical galaxies, which tend to have a circular shape. Accordingly, if one of the S\'ersic indexes required by the user is $n=4$, there is a boolean parameter which indicates whether these galaxies will have only a circular shape, or an elliptical shape (which allows different inclination and eccentricity values). 
Figure~\ref{galaxy_stamps} shows examples of simulated galaxies with different features.

For each redshift bin, we create a set of stamps each representing an artificial galaxy with total flux given by equation~\ref{eq2}. The value of the flux in each individual pixel at position $(x_i,y_i)$ and size $\Delta r$ is calculated numerically by integrating the surface brightness profile:  

\begin{equation}
    L(x_i,y_i) = \int_{x_i-\Delta r/2}^{x_i+ \Delta r/2} \int_{y_i -\Delta r/2}^{y_i + \Delta_r/2} I(r) dx dy
\end{equation}

where $r^2=(x^2+y^2)$.

We note that previous approaches to completeness simulations have resorted to oversampling the inner pixels of the artificial sources as a balance between accuracy and computational speedup \citep[e.g.:][]{peng2002, haussler2007}. However, as \texttt{GLACiAR} is tailored for high redshift galaxies, which are typically marginally resolved, we prefer to implement a highly accurate calculation of the flux in each individual pixel. 

The artificial sources generated by \texttt{GLACiAR} do not include Poisson noise. This is motivated by the fact that Poisson noise becomes dominant over other components (background, readout, and dark current noise) only in a regime where the source is detected at high confidence ($S/N\gtrsim 50$). Under these conditions, completeness simulations are not required. For example, we verified from the Hubble Space Telescope WFC3 Exposure Time Calculator that for a compact source in the F160W filter, Poisson noise becomes greater than the sky background at $S/N>80$. 

For each set of parameters, subsets with all the possible galaxies in terms of inclination and eccentricity for each S\'ersic index are generated. All the simulated galaxies in each subset have the same redshift, meaning that the parameters that change, apart from $n$ are the slope $\beta$, the input magnitude, eccentricity, and inclination angle. Both $\beta$ and the input magnitude only modify the flux, i.e. the shape of the surface brightness profile of the simulated galaxy remains the same except for a scaling factor. Hence, we do not need to recalculate the flux in each pixel for these galaxies as we can just apply a global re-scaling. In the case of the eccentricity and inclination angle, these parameters change the shape of the source and distribution of its flux. Given that, we generate all possible combinations for each subset with the same S\'ersic index and redshift. Note that the redshift also changes the distribution of the flux as we define $R_{eff}$ as a function of $z$.

\subsection{Point Spread Function Convolution}\label{psf}

The PSF describes the imaging system response to a point input, and we take it into account to properly include the instrumental response into our model mock galaxies. In order to do this, we need a user-supplied PSF image, which is convolved with the artificial galaxy images through the  the \texttt{python} module \texttt{convolution.convolve} from \texttt{Astropy}. For commonly-used \textit{HST} filters, we already include Tiny Tim PSF data\footnote{http://www.stsci.edu/hst/observatory/focus/TinyTim} in the `$psf$' folder. If the user desires to apply \texttt{GLACiAR} to filters not listed in the code, the corresponding files can be added to that folder.

\subsection{Positions}

After generating the simulated galaxy stamps, their position $(x,y)$ is assigned within the science image. These coordinates $(x,y)$ correspond to the pixel where the center of the stamp will be placed, and are generated as pairs of uniform random numbers across the pixel range in the science image. Two conditions are required to accept the pair: First, for physical reasons, a simulated galaxy cannot be blended with another simulated galaxy (but no limitation is imposed to blending with sources in the original science image) and second, the center of the simulated source must fall inside the science image boundaries (technically implemented by requiring the pixel to have a value different from zero in the science image). 
The artificial source positions generated are saved for comparison in the subsequent steps.

\subsection{Multi-band data}

\texttt{GLACiAR} is structured to handle multiple, user-specified photometric bands. Depending on the redshift of the simulated source and the slope of its spectrum $\beta$, synthetic images will have different magnitudes in different bands. To calculate them, the code starts from the spectrum defined in Equation~\ref{eq_spectrum} (see Figure~\ref{spectrum} for an example), and it convolves it with the relevant filter transmission curve using the function \texttt{pysysp} from the package \texttt{PyPI}. Input files for a set of default HST filters are included in our release. If the user requires a different filter that is not part of the \texttt{GLACiAR}'s sample, they can add it by adding the transmission file in the folder `filters'. 

After calculating the flux of the simulated source in each filter, the postage stamp image of the artificial galaxy is rescaled to that total flux. In order to save time, we let all the simulated galaxies in a single recovery simulation iteration have the same value of $\beta$, so there is no need to repeat the filter convolution for sources at the same redshift, and sample instead a different value of $\beta$ in each iteration. This saves computational resources without impacting the end results since (1) we employ a sufficient number of iterations ($n_{iter}=100$ by default) to sample the $\beta$ distribution reasonably well, and (2) changes in $\beta$ produce only relatively small differences in colours ($\Delta m < 0.1$) for default input choices. Therefore it's not necessary to sample a different $\beta$ value for each galaxy.

Finally, the artificial galaxies stamps are added to the science images in the corresponding bands, if their total magnitude in that band is below a critical threshold ($m_{AB}\leq 50$ by default). 

\subsection{Source Identification}\label{recovery}

We run a source identification software (\texttt{SExtractor} in this case) on the original images, as well as on the new images with the simulated galaxies, to create source catalogs. In order to do that, the user must provide a configuration file under the folder `SExtractor\_files'. If no file is provided, the software uses the default one, `parameters.sextractor'. The filter file also needs to be copied here. We provide one example with the filter `gauss\_2.0\_5x5.conv'.


\texttt{GLACiAR} calls \texttt{SExtractor} to run over all the science images with added artificial sources generated in each iteration; it produces new catalogs and new segmentation maps for each of them. To ease storage space requirements, segmentation maps are deleted after use by default. 

To study the recovery fraction, the segmentation map of the original image is compared with the segmentation map of the image containing the simulated galaxies. The positions where the simulated galaxies were placed have been recorded, therefore the new segmentation map values in that position can be checked. It is possible that the new source is not found by \texttt{SExtractor} in the actual position that was placed in, thus we allow a certain margin, examining the values of the new segmentation map over a grid of 3x3 pixels centered in the original input position. If any of the values of this grid is not zero, the ID number of the object that is there is saved (i.e. the value of that pixel in the segmentation map). To determine whether that object is blended, we check in the original segmentation map the values of the pixels where the simulated object lies. If any of the pixel values are different from zero, the object is flagged as blended. If the real source blended with the simulated galaxy has an original magnitude fainter than the simulated galaxy input magnitude, we still consider the simulated object successfully recovered. On the other hand, if the original science source is brighter, an extra test is performed. If less than 25\% of the pixels of the new object overlap with the original object, and there is a difference smaller than 25\% between recovered and input flux of the simulated object, we still consider it as recovered, while if any of these two requirements are not met, we flag the artificial source as not recovered. This is a conservative (and moderately computationally intensive) approach on assessing blending, but it has advantages of taking into full account the arbitrary shape of foreground sources and the extent of the overlap of the segmentation maps when compared to a distance-based approach. We also note that 25\% overlap is an arbitrary threshold that we fined-tuned based on experimentation, which users are free to modify.

\begin{figure*}
    \centering
    \includegraphics[scale=0.3]{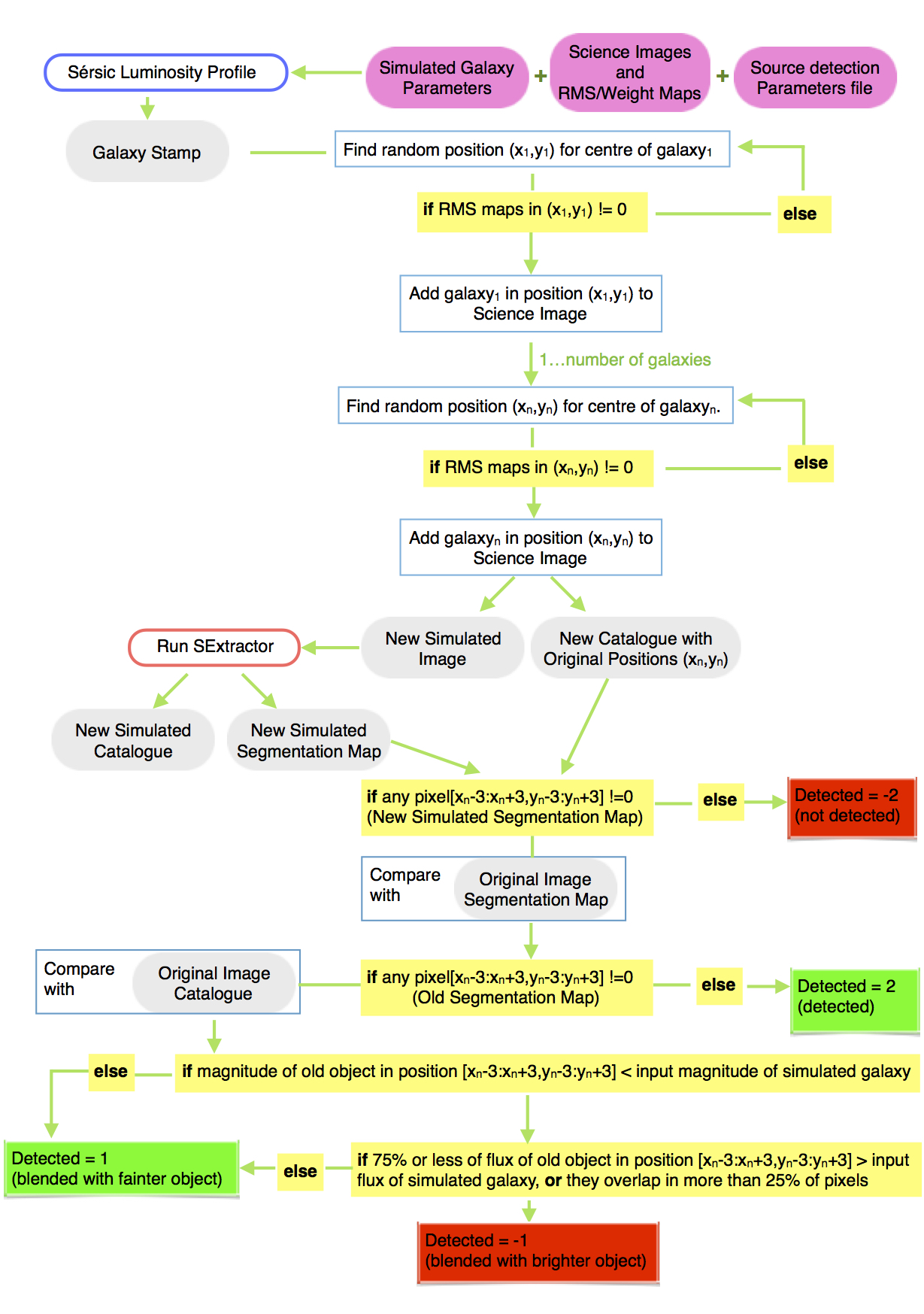}
    \caption{Diagram with a detailed explanation of how \texttt{GLACiAR}'s algorithm structure, focusing in particular on the blending classification.}
    \label{fig:diagram_detailed}
\end{figure*}

To summarize this process, Figure~\ref{fig:diagram_detailed} shows a flow chart with a detailed explanation of, in particular, the blending and recovering of sources. Furthermore, Figure~\ref{maps} shows an example of the identification of the simulated galaxies in one of the fields of the BoRG survey.

\subsection{Multi-band photometric output}

The ultimate output of \texttt{GLACiAR} is a multi-band photometric catalog that lists input and output properties of the artificial objects, including a flag to indicate whether entries have been marked as successfully recovered. This catalog naturally allows the user to run a customized data analysis to measure completeness and source recovery using the same criteria that the user would apply to actual science data (whether a dropout technique or a photometric redshift estimation is desired). For convenience of Lyman-break selection users, \texttt{GLACiAR} has a module that performs statistical analysis of the recovery as a function of input redshift and magnitude.

\section{IMPLEMENTING AND RUNNING THE CODE}\label{implementation}

The code is in the \texttt{Github} repository https://github.com/danielacarrasco/GLACiAR. The user should download the code, change the input parameters, and add any files if needed. Detailed instructions are provided in a README file. A brief description of the parameters and required files follows below.

\subsection{Input Parameters}\label{input_parameters}

The parameters needed to run the simulation are found in the file `\textit{parameters.yaml}'. Some of them need to be specified by the user, while others can be either inputted or left blank, in which case they take a default value. A description of all the parameters is given in Appendix \ref{appendix}.

\begin{figure*}
    \centering
    \includegraphics[scale=0.3]{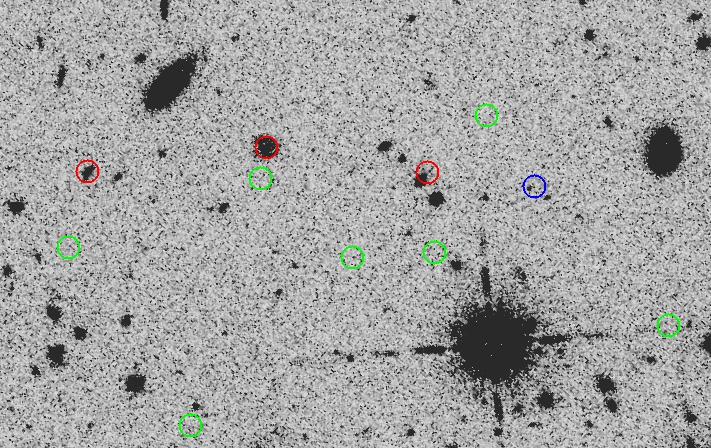}
    \includegraphics[scale=0.3]{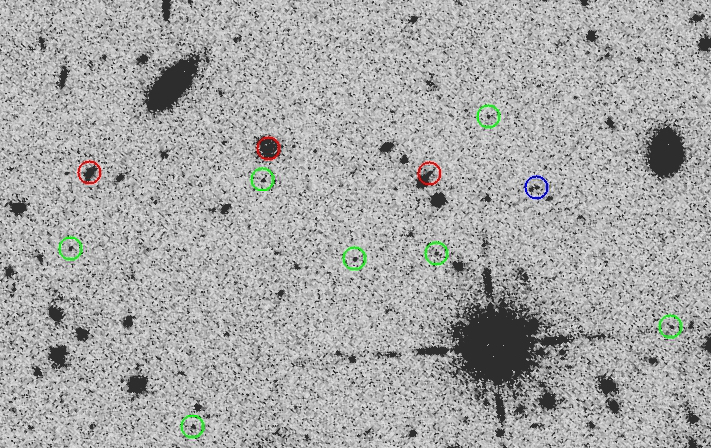}\\
    \includegraphics[scale=0.3]{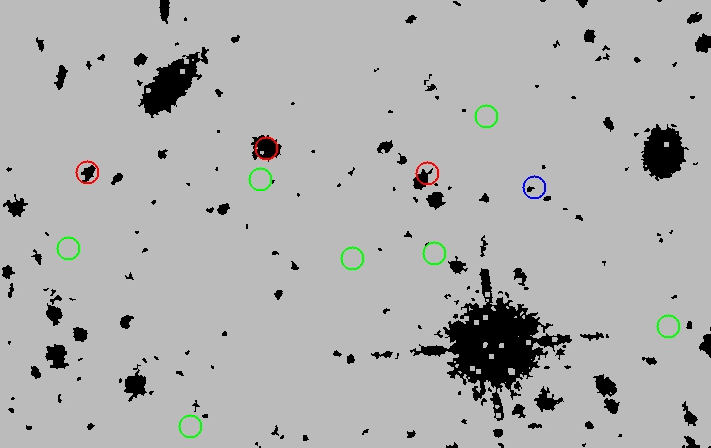}
    \includegraphics[scale=0.3]{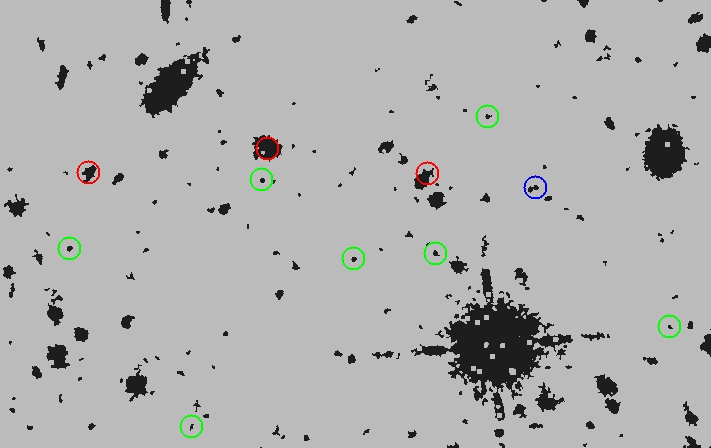}
    \caption{Illustration of \texttt{GLACiAR}'s application to BoRG field $borg\_$0835+2456. textit{Top left:} Original science image.\textit{Top right:} Science image plus simulated galaxies with an input magnitude of $m_{H}=26.0$ indicated by colored circles. \textit{Bottom left:} \texttt{SExtractor} Segmentation map for the original science image. \textit{Bottom right:} Segmentation map after running \texttt{SExtractor} on the image that includes simulated galaxies. The color of the circles encodes detection of the simulated sources with green indicating recovery for an isolated galaxy, blue recovery but source blended with a fainter object. Detection failures are shown in red.}
    \label{maps}
\end{figure*}

\subsection{Required Files}\label{files_needed}

The files required for the algorithm are described below. More details on their format and location can be found in the README file on \texttt{Github}.

\begin{itemize}
\item[] \textsf{Science images:} All the images in which the simulation is going to be run on. It must include all the different fields and bands as well.
\item[] \textsf{List:} Text file with the names of the fields from the survey. This list is given as an input parameter (see section~\ref{input_parameters}). 
\item[] \textsf{\texttt{SExtractor} parameters:} As discussed in section~\ref{recovery}, \texttt{GLACiAR} invokes instances of \texttt{SExtractor} on the images (original and with simulated galaxies). To run that external software, a file defining the parameters is needed. There is an example provided under the folder `SExtractor\_files' (based on the BoRG survey source detection pipeline), which will be used if no other file is provided, but we recommend the user to customize this input to optimize their specific analysis.
\item[] \textsf{RMS maps or weight maps:} Frames having the same size as the science image that describe the noise intensity at each pixel. They are necessary only if required for the \texttt{SExtractor} parameters. They are defined as:
\begin{equation}
    weight = \frac{1}{variance} = \frac{1}{rms^{2}}
\end{equation}
\item[] \textsf{PSF}: PSF data for filters/instruments not currently included in the release can be added in this folder by the user (see~\ref{psf} for more details).
\end{itemize}

\section{EXAMPLE APPLICATION}\label{results}

To illustrate \texttt{GLACiAR}'s use, we apply it to estimate the completeness and source recovery of a large HST imaging program, the Brightest of Reionizing Galaxies Survey (BoRG), focused on identifying $L>L_*$ galaxies at $z\gtrsim 8$ along random lines of sights \citep{trenti2011,trenti2012,bradley2012,schmidt2014,calvi2016,bernard}. Specifically, we focus on characterizing the J-dropout source recovery (galaxies at $z\sim 10$) and compare our results with those in \citet{bernard}. The results are discussed throughout this section, and they can be seen in Figure~\ref{fig:completeness_borg} and~\ref{fig:dropouts_borg}.

\subsubsection{Data}\label{borg}

The dataset considered here is the BoRG[z8] subset, consisting of core BoRG pointings (GO11700, 12572, 12905), augmented by other pure parallel archival data (GO 11702, PI Yan, \citet{yan2011}) and COS GTO coordinanted parallel observations. For a detailed description of the survey, we refer to \citet{trenti2011}; \citet{bradley2012}; \citet{schmidt2014}. We use the 2014 (DR3) public release of the data.\footnote{https://archive.stsci.edu/prepds/borgy/}, which consists of 71 independent pointings covering a total area of $\sim350$ arcmin$^{2}$. All fields were imaged in 4 bands: F098M ($Y_{098}$), F125W ($J_{125}$), F160W ($H_{160}$), and an optical band F606W ($V_{606}$) or ($V_{600}$). The BoRG[z8] public data release consists of reduced and aligned science images produced with \texttt{MultiDrizzle} \citep{multidrizzle}, a pixel scale of 0.08, and associated weight maps \citep{bradley2012,schmidt2014}. The 5$\sigma$ limiting magnitudes for point sources and aperture $r=0.2''$ vary between $m_{AB}=25.6-27.4$, with a typical value of $m_{AB}\sim26.7$.


\subsection{Redshift Selection/Dropouts criteria}

We use \texttt{GLACiAR} for recovery of simulated sources in the redshift range of $z\sim 10$. In order to do this, we apply a selection criteria to find $J_{125}$ dropouts following \citet{bernard}:
\begin{itemize}
\item[-]$S/N_{160}\geq 8.0$
\item[-]$S/N_{V}<1.5$
\item[-]$S/N_{098}<1.5$
\item[-]$J_{125}-H_{160}>1.5$
\end{itemize}
Note that while these criteria are set as default in the code, their selection is fully customizable by the user. 

\subsection{Completeness and Source Recovery Output}

The main results produced by the program can be summarized in three tables described below, including an example for the first two (see Tables~\ref{ResultsStats} and~\ref{ResultsGalaxiesBand}).

First, the statistics of what fraction of the galaxies placed in the image were identified and how many were recovered at the corresponding redshift with the selection technique. Table~\ref{ResultsStats} shows an example of its structure for our BoRG dataset.

\begin{table*}
    \centering
    \begin{tabular}{cccccccccccc}
        \hline \hline
        z$^{a}$ & m$^{b}$ & N\_Obj$^{c}$ & $S=0^{d}$ & $S=1,2^{e}$ & $S=-1^{f}$ & $S=-2^{g}$ & $S=-3^{h}$ & N\_Rec$^{i}$ & N\_Drop$^{j}$ & Rec$^{k}$ & Drops$^{l}$\\ \hline
        9.0 & 24.1 & 300 & 218 & 26 & 50 & 4 & 2 & 268 & 0 & 0.89 & 0.0 \\ 
        9.0 & 24.3 & 1000 & 751 & 62 & 169 & 13 & 5 & 920 & 0 & 0.92 & 0.0\\
        9.0 & 24.5 & 1500 & 1112 & 94 & 257 & 26 & 11 & 1369 & 0  & 0.91 & 0.0\\
        \vdots & \vdots & \vdots & \vdots & \vdots & \vdots & \vdots & \vdots & \vdots & \vdots & \vdots\\
        10.0 & 24.1 & 300 & 211 & 17 & 63 & 5 & 4 & 274 & 101 & 0.91 & 0.34\\
        \vdots & \vdots & \vdots & \vdots & \vdots & \vdots & \vdots & \vdots & \vdots & \vdots & \vdots\\
        11.8 & 27.9 & 600 & 0 & 72 & 0 & 34 & 494 & 0 & 0 & 0.0 & 0.0\\\hline
    \end{tabular}
    \caption{Example of the file produced by the simulation with the statistics for each redshift and magnitude.}
    \tabnote{$^{a}$ Input redshift of the simulated galaxy.}
    \tabnote{$^{b}$ Magnitude bin that represents the median value of the bins.}
    \tabnote{$^{c}$ Number of objects inputted for that redshift and magnitude bin in all the iterations.}
    \tabnote{$^{d}$ Number of galaxies recovered by \texttt{SExtractor} that were isolated.}
    \tabnote{$^{e}$ Number of artificial sources recovered that were blended with a fainter object.}
    \tabnote{$^{f}$ Number of artificial sources recovered that were blended with a brighter object.}
    \tabnote{$^{g}$ Number of artificial sources that were detected by \texttt{SExtractor} but with a $S/N$ under the required threshold.}
    \tabnote{$^{h}$ Number of artificial sources that were not detected by \texttt{SExtractor}.}
    \tabnote{$^{i}$ Number of recovered artificial sources: $(d+e)$.}
    \tabnote{$^{j}$ Number of artificial sources that passed the dropout selection criteria}
    \tabnote{$^{k}$ Fraction of not recovered artificial sources : $\frac{i}{c}$.}
    \tabnote{$^{l}$ Fraction of artificial sources that passed the selection criteria$\frac{j}{c}$.}
    \label{ResultsStats}
\end{table*}

Second, a table with more detail about the galaxies that were inserted and the recovering results, several tables (one for each redshift step) are produced with all the galaxies that were placed in the simulations at that redshift. They have the recovered magnitude in the detection band, the identification status, the ID given by \texttt{SExtractor}, among others. The structure is shown in Table~\ref{ResultsGalaxiesBand}

\begin{table*}
    \centering
    \begin{tabular}{cccccc}
        \hline \hline
        Initial Mag$^{a}$ & iteration$^{b}$ & ID Number$^{c}$ & Input Magnitude$^{d}$ & Output Magnitude$^{e}$ & Identification Status$^{f}$\\ \hline
        24.1 & 1 & 319 & 25.922 & 26.255 & 0\\
        24.1 & 1 & 213 & 25.922 & 26.088 & 0\\
        \vdots & \vdots & \vdots & \vdots & \vdots & \vdots\\
        27.9 & 10 & 39 & 26.952 & 23.627 & -1\\
        27.9 & 10 & 0 & 26.952 & -99.000 & -3\\\hline
    \end{tabular}
    \caption{Example of the file produced by the \texttt{GLACiAR} with information of all the simulated galaxies.}
    \tabnote{$^{a}$ Magnitude corresponding to the input flux for the star. This is not the same as $^{d}$ as the input magnitude changes depending on the $\beta$ value and size of the object.}
    \tabnote{$^{b}$ Iteration number.}
    \tabnote{$^{c}$ Identification number given by \texttt{SExtractor} after it runs on the image with the simulated galaxies. This number is unique for every iteration for a given magnitude and redshift.}
    \tabnote{$^{d}$ Magnitude corresponding to the added flux inside all the pixels that the source includes.}
    \tabnote{$^{e}$ Magnitude of the source found with \texttt{SExtractor} after it runs on the image with the simulated galaxies.}
    \tabnote{$^{f}$ Integer number that indicates whether a source has been recovered and/or is blended.}
    \label{ResultsGalaxiesBand}
\end{table*}

Third, one last table, which is useful for redshift selection. Given that the number of bands is variable, and it can be large, this table is released in a Python-specific compact binary representation (using the \texttt{pickle} module). It contains the ID of the object, input magnitude, status, magnitudes in all bands, and $S/N$ for each band as well.

%

\subsection{Results and Comparisons}

\begin{figure}[ht!]
\centering
    \includegraphics[scale=.75]{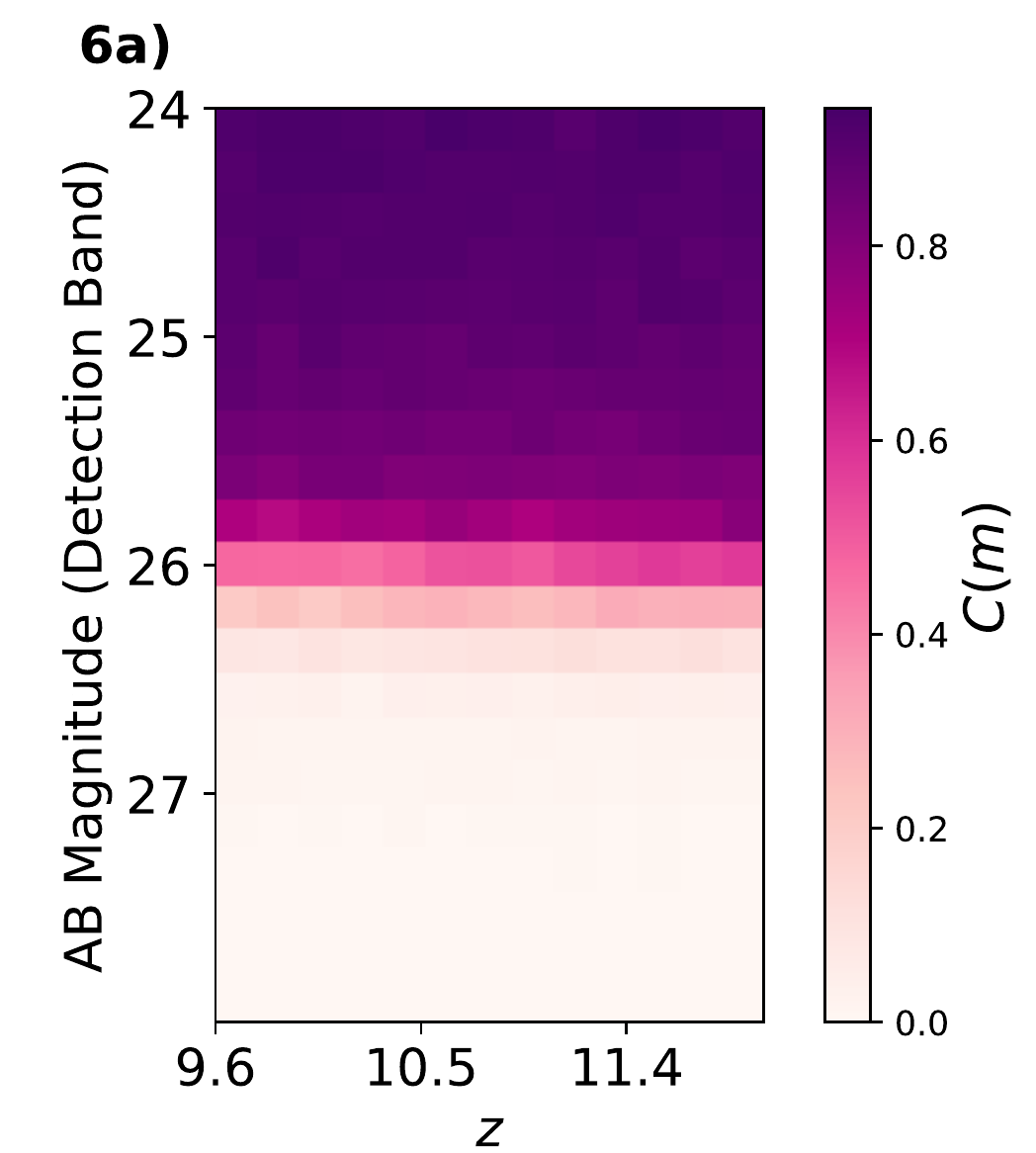}
    \includegraphics[scale=.58]{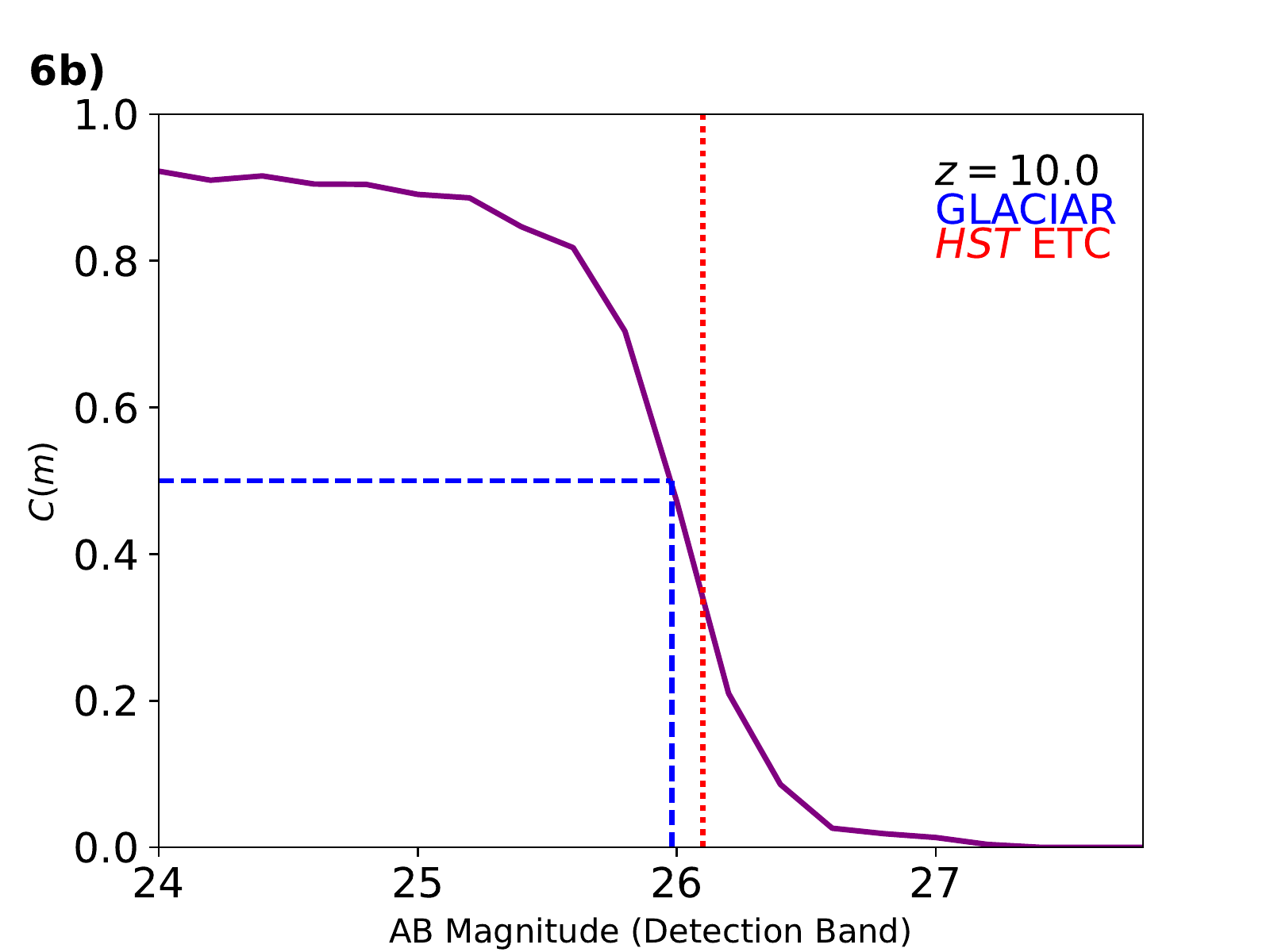}  
    \caption{Completeness selection plots produced by our simulation for the BoRG field $borg\_$0440-5244 in F160W. The top panel shows the completeness for a range of redshifts $z=9.6-12.0$, and the bottom panel shows a slice of those results for $z=10$. The completeness is around $\sim90\%$ up to $m_{AB}\sim25.0$, and it drops to $0.0\%$ for $m_{AB}\gtrsim27.0$. The blue dashed line shows the $50\%$ calculated by \texttt{GLACiAR} ($m_{AB}=25.98$). The red dashed line shows the limiting magnitude at which a point source with circular radius of 0.2'' and a  spectrum following a power law $F(\lambda) = \lambda^{-1}$ is detected at a $S/N = 8$ according to the \textit{HST} exposure time calculator ($m_{AB}=26.10$).}
    \label{fig:completeness_borg}
\end{figure}

We run the simulation for the whole BoRG[z8] survey. As an example, the results for one field ($borg\_$0440-5244) are shown in Figure~\ref{fig:completeness_borg}; Figure 6a shows the completeness fraction $C(m)$ for different redshifts as a function of the input magnitude, while Figure 6b is a slice of $C(m)$ at a fixed redshift ($z=10.0$). As we can see, the completeness is around $C(m)\sim90\%$ up to a magnitude of $m_{AB}\sim25.0$, and it drops to $C(m)=0.0\%$ for $m_{AB}\gtrsim27.1$, while at $m_{AB}\sim25.98$, we find a completeness of $C(m)=50\%$. This is expected from when comparing with the results from \textit{HST} exposure time calculator\footnote{http://etc.stsci.edu/etc/input/wfc3ir/imaging/}: a galaxy at $z=10.0$ in an image with the characteristics of the field we are running our simulations on, gives as a $S/N$ ratio of $\sim8.0$ at a magnitude of $m_{AB}=26.1$ in the $H_{160}$ band for a point source galaxy with circular radius of 0.2'' and a power law $F(\lambda) = \lambda^{-1}$ spectrum. 

\begin{figure}[ht!]
    \centering
    \includegraphics[scale=.8]{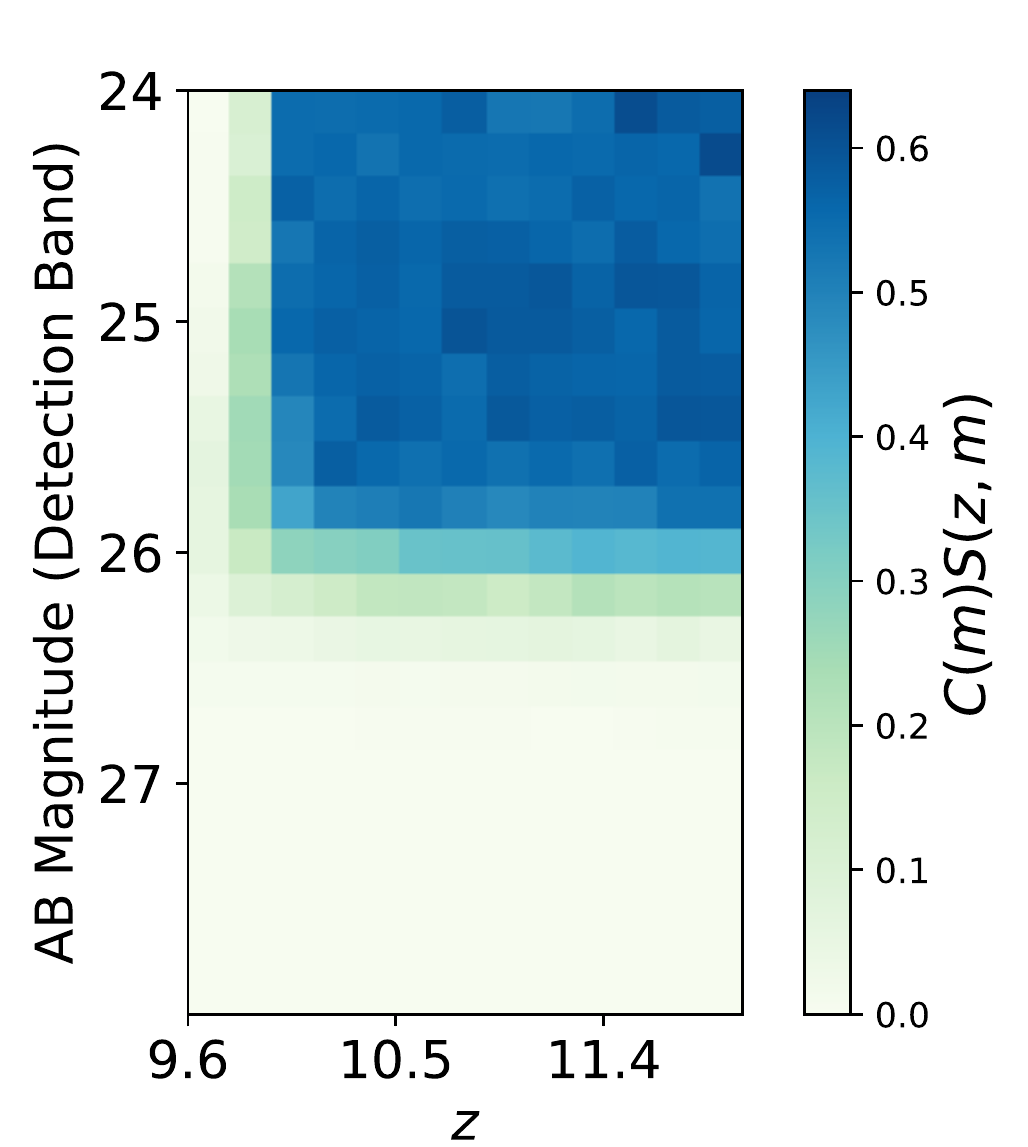}
    \includegraphics[scale=.8]{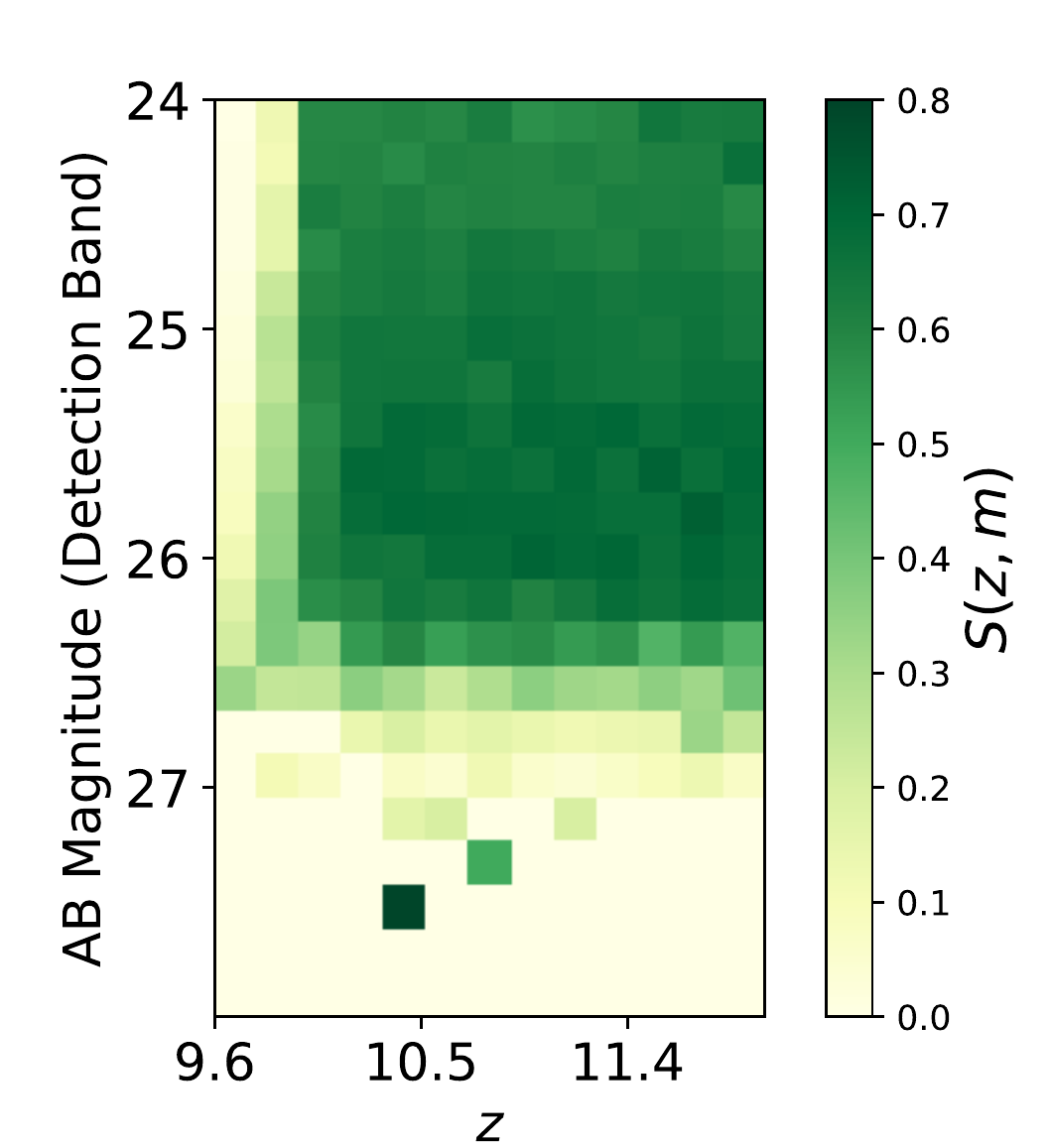}
    \caption{Dropouts selection plots produced by our code for the BoRG field $borg\_$0440-5244 for redshift $z\sim10$. The top panel shows the dropouts found from all the galaxies inserted ($C(m)S(z,m)$), while the bottom panel shows the fraction of recovered dropouts ($S(z,m)$) for artificial sources that are successfully identified in the detection band. Note that the bottom panel becomes noisy for $m_{AB}>27.0$ since $S(z,m)$ is computed only using the small number of faint artificial galaxies that are identified with success. The top panel does not suffer from such noise, instead.}
    \label{fig:dropouts_borg}
\end{figure}

The results of the dropout selection for the same field are shown in Figure~\ref{fig:dropouts_borg}. We can compare our results with the ones from \citet{bernard} (bottom panel of Figure 4 in their paper), where we can see the selection function $C(m)S(z,m)$ for the field $borg\_$0440-5244. Our results achieve a maximum of $\sim64\%$ recovery, to be compared against the maximum $\sim75\%$ recovery reported in their paper. As we have full access to the code used to produce both sets of results, we can attempt to understand the origin of this discrepancy. First of all, there is a difference in $C(m)$ in the range of $m_{AB} = 25.5 - 26.0$, that is most likely attributed to the definition of successful recovery for blended or potentially blended sources. In fact, when comparing the results for recovery of isolated objects \texttt{GLACiAR} obtains the same results. The completeness analysis in \citet{bernard} considers sources as blended based on the distance from the center of the objects, i.e. if the detected object is closer than a certain distance (in pixels) from the center of an object in the original science catalog, then it classifies the artifical source as blended. In this respect, \texttt{GLACiAR} improves upon the previous analysis by carrying out a more sophisticated analysis based on comparison of the segmentation maps, which take into account the actual spatial extension of the sources, instead of limiting the analysis to catalog output. 

Another key difference originates from how our galaxies are simulated: we simulate images in all the bands, even when the expected is negligible given the spectrum of the artifical source. In the case of \citet{bernard}, the V-band ($F600LP$ or $F606W$) non-detection requirement was not simulated since it was assumed that artificial sources had no flux in that band. To account for this, the selection function computed excluding the V-band non-detection requirement was reduced by $6.2\%$, which derives from the assumption that the S/N distribution in the V-band photometry would follow Gaussian statistics. \texttt{GLACiAR} performs instead a full color simulation and our results indicate that non-Gaussian tails contribute to exclude a larger fraction of objects at bright magnitudes. Indeed, if we replicate the approach by \citet{bernard} we obtain instead results consistent with that study (for isolated sources). Thus all differences are understood and the comparison contributes to validate the accuracy of \texttt{GLACiAR}.

Note that \texttt{GLACiAR} results for $C(m)$ and $S(z,m)$ are provided as a function of the intrinsic magnitude of the simulated images. Previous studies, including \citet{bernard}, may present completeness as a function of recovered output magnitude instead. Since in the latter case a specific LF for the simulated sources has to be assumed to map intrinsic to observed completeness through a transfer function, we opted to setup the output of \texttt{GLACiAR} to provide only the fundamental quantity, and leave derivation of an observed completeness to the user if needed.

\section{DISCUSSION AND SUMMARY}\label{discussion}

In this paper, we present a new tool to estimate the completeness in galaxy surveys, \texttt{GLACiAR}. This algorithm creates an artificial galaxy stamp that follows a S\'ersic profile with parameters such as the size, S\'ersic index, input magnitude, input redshift, filters, among others, that are chosen by the user. After creating the galaxies, they are added to the science image. A source identification algorithm is run on the science images and on the images with the simulated galaxies, in order to study the recovery of these mock galaxies.
After the source catalogs are produced, we match the newly found objects with the positions in which the simulated galaxies were originally inserted, and we cross-match the area of the segmentation maps corresponding to these new sources with the ones from the original catalogs, so the status of these galaxies can be determined. These statuses can be categorized in four groups: detected and isolated, blended with fainter object, blended with brighter object, and not detected. If a source falls into one of the two first categories only, it is considered detected (an example can be seen in Figure~\ref{maps}). The final product of the algorithm are three types of tables, with the information of the statistics about the recovery, the detected galaxies, and all the galaxies.
To illustrate the use of the new tool, and to validate it against previous literature analysis, we applied \texttt{GLACiAR} to analysis of the selection function for $z\sim 10$ galaxies in the BoRG[z8] survey, comparing our results to the recent work by \citet{bernard}. Section~\ref{results} discusses the comparison in detail, with the key summary being that while (minor) differences are present, these can be attributed to improvements introduced by \texttt{GLACiAR} and are fully understood. In particular, the improved completeness analysis is more realistic in its treatment of non-Gaussian noise for all survey bands, and includes a sophisticated comparison between segmentation maps to identify blended objects to high reliability.


This initial application demonstrates that \texttt{GLACiAR} is a valuable tool to unify the completeness estimation in galaxy surveys. So far, the code is limited to surveys where the detection of the sources is done by \texttt{SExtractor}, but its structure has been designed to allow a future upgrade of capabilities by inclusion of \texttt{photutils} as well.
More broadly, the code is flexible allowing, for instance, the possibility of modifying the redshift selection criteria along with the fraction of galaxies that follow different values of $n$ for the S\'ersic luminosity profile. This makes \texttt{GLACiAR} suitable for a range of different applications in galaxy formation and evolution observations, including studies of LFs, contamination rates in galaxy surveys, characteristics of selected galaxies in redshift selections, among others. A future release of the code will also incorporate a module to account for weak and strong lensing magnification maps, with applications to galaxy cluster surveys such as the \textit{Frontier Fields} initiative.

\begin{acknowledgements}
We thank the referee for helpful suggestions and comments that improved the paper. This work was partially supported by grants HST/GO 13767, 12905, and 12572, and by the ARC Centre of Excellence for All-Sky Astrophysics in 3 Dimensions (ASTRO-3D). 
\end{acknowledgements}

\begin{appendix}

\section{Description of input parameters}\label{appendix}

Below there is a list with a brief description of all the parameters used to run \texttt{GLACiAR}:

\begin{itemize}
\item[] \textsf{n\_galaxies:} Number of galaxies per image to place in each iteration ($\texttt{default} = 100$).
\item[] \textsf{n\_iterations:} Number of iterations, i.e. the number of times the simulation is going to be run on each image for galaxies with the same redshift and magnitude ($\texttt{default} = 100$).
\item[] \textsf{mag\_bins:} The number of desired magnitude bins. For a simulation run from $m_{1} = 24.0$ to $m_{2} = 25.0$ in steps of 0.2 magnitudes, there will be 6 bins ($\texttt{default} = 20$).
\item[] \textsf{min\_mag:} Brightest magnitude of the simulated galaxies ($\texttt{default} = 24.1$).
\item[] \textsf{max\_mag:} Faintest magnitude of the simulated galaxies ($\texttt{default} = 27.9$).
\item[] \textsf{z\_bins:} The number of desired redshift bins. For a simulation run from $z_{1} = 9.5$ to $z_{2} = 10.5$ in steps of 0.2, there will be 6 bins ($\texttt{default} = 15$).
\item[] \textsf{min\_z:} Minimum redshift of the simulated galaxies ($\texttt{default} = 9.0$).
\item[] \textsf{max\_z:} Maximum redshift of the simulated galaxies ($\texttt{default} = 11.9$).
\item[] \textsf{n\_bands:} Number of filters the survey images have been observed in. If not specified, it will raise an error.
\item[] \textsf{detection\_band:} Band in which objects are identified. If not specified, it will raise an error.
\item[] \textsf{lambda\_detection:} Central wavelength in angstroms of the detection band. If not specified, it will raise an error.
\item[] \textsf{bands:} Name of the bands from \textsf{n\_bands}. The detection band has to be the first entry in the list. If not specified, it will raise an error.
\item[] \textsf{zeropoints:} Zeropoint values corresponding to each band. The entries must follow the same order as \textsf{bands}. Default values are set as 25.0. 
\item[] \textsf{gain\_values:} Gain values for each band. The entries must follow the same order as \textsf{bands}. If not specified, it will raise an error.
\item[] \textsf{list\_of\_fields:} Text file containing a list with the names of the fields the simulation will run for, which can be one or more. If not specified, it will raise an error.
\item[] \textsf{R\_eff:} Effective radius in kpc for a simulated galaxy at $z=6$. It is the half light radius, i.e. the radius within half of the light emitted is enclosed. This value changes with the redshift as $(1+z)^{-1}$ ($\texttt{default} = 1.075$ kpc).
\item[] \textsf{beta\_mean:} Mean value for a Gaussian distribution of the UV spectral slope (Section~\ref{artificialgalaxy}). ($\texttt{default} = -2.2$).
\item[] \textsf{beta\_sd:} Standard deviation for the for a Gaussian distribution of the slope of the spectrum as explained in  Section~\ref{artificialgalaxy}. ($\texttt{default} = 0.4$).
\item[] \textsf{size\_pix:} Pixel scale for the images in arcsec ($\texttt{default} = 0.08$). 
\item[] \textsf{path\_to\_images:} Directory where the images are located. The program will create a folder inside it with the results. If not specified, it will raise an error.
\item[] \textsf{image\_name:} Name of the images. They all should have the same name with the name of the field (list\_of\_fields) and band written at the end, as follows: `image$\_$name+field+band.fits'. If not specified, it will raise an error.
\item[] \textsf{types\_galaxies:} Number indicating the amount of S\'ersic indexes. ($\texttt{default} = 2$). 
\item[] \textsf{sersic\_indexes:} Value of the S\'ersic index parameter $n$ for the number of \textsf{types\_galaxies} ($\texttt{default} = [1,4]$). 
\item[] \textsf{fraction\_type\_galaxies:} Fraction of galaxies corresponding the the S\'ersic indexes given ($\texttt{default} = [0.5,0.5]$).
\item[] \textsf{ibins:} Number of bins for the inclination angle. The inclinations can vary from $0^{\circ}$ to $90^{\circ}$, i.e., if 10 bins are chosen, the variations will be of $9^{\circ}$. One bin indicates no variation of inclination angle. ($\texttt{default} = 1$).
\item[] \textsf{ebins:}  Number of bins for the eccentricity. The values can vary 0 to 1, i.e., if 10 bins are chosen, the variations will be of 0.1. One bin indicates only circular shapes ($\texttt{default} = 1$).
\item[] \textsf{min\_sn:} Minimum $S/N$ ratio in the detection band for an object to be considered detected by \texttt{SExtractor}. ($\texttt{default} = 8.0$)
\item[] \textsf{dropouts:} Boolean that indicates whether the user desires to run a dropout selection ($\texttt{default} = False$).
\item[] \textsf{de\_Vacouleur:} Boolean that indicates whether the user wants to make an exemption for de Vaucouleur galaxies. If true, galaxies with $n=4$ will only have circular shape ($\texttt{default} = False$).
\end{itemize}

\end{appendix}

\nocite*{}
\bibliographystyle{pasa-mnras}
\bibliography{biblio}

\begin{thebibliography}{}
\makeatletter
\relax
\def\mn@urlcharsother{\let\do\@makeother \do\$\do\&\do\#\do\^\do\_\do\%\do\~}
\definecolor{darkblue}{rgb}{0,0,0.597656}
\def\mndoi{\begingroup\mn@urlcharsother \@ifnextchar [ {\mndoi@} {\mndoi@[]}}
\def\mndoi@[#1]#2{\def\@tempa{#1}\ifx\@tempa\@empty \href
  {http://dx.doi.org/#2} {\textcolor{darkblue}{doi:#2}}\else \href
  {http://dx.doi.org/#2} {\textcolor{darkblue}{#1}}\fi \endgroup}
\def\mn@eprint#1#2{\mn@eprint@#1:#2::\@nil}
\def\mn@eprint@arXiv#1{\href {http://arxiv.org/abs/#1} {{\tt arXiv:#1}}}
\def\mn@eprint@dblp#1{\href {http://dblp.uni-trier.de/rec/bibtex/#1.xml}
  {dblp:#1}}
\def\mn@eprint@#1:#2:#3:#4\@nil{\def\@tempa {#1}\def\@tempb {#2}\def\@tempc
  {#3}\ifx \@tempc \@empty \let \@tempc \@tempb \let \@tempb \@tempa \fi \ifx
  \@tempb \@empty \def\@tempb {arXiv}\fi \@ifundefined
  {mn@eprint@\@tempb}{\@tempb:\@tempc}{\expandafter \expandafter \csname
  mn@eprint@\@tempb\endcsname \expandafter{\@tempc}}}

\bibitem[\protect\citeauthoryear{{Atek} et~al.,}{{Atek}
  et~al.}{2015}]{atek2015}
{Atek} H.,  et~al., 2015, \mndoi [\apj] {10.1088/0004-637X/800/1/18}, \href
  {http://adsabs.harvard.edu/abs/2015ApJ...800...18A} {800, 18}

\bibitem[\protect\citeauthoryear{{Beckwith} et~al.,}{{Beckwith}
  et~al.}{2006}]{hudf}
{Beckwith} S.~V.~W.,  et~al., 2006, \mndoi [\aj] {10.1086/507302}, \href
  {http://adsabs.harvard.edu/abs/2006AJ....132.1729B} {132, 1729}

\bibitem[\protect\citeauthoryear{{Bernard} et~al.,}{{Bernard}
  et~al.}{2016}]{bernard}
{Bernard} S.~R.,  et~al., 2016, \mndoi [\apj] {10.3847/0004-637X/827/1/76},
  \href {http://adsabs.harvard.edu/abs/2016ApJ...827...76B} {827, 76}

\bibitem[\protect\citeauthoryear{{Bershady}, {Lowenthal}  \& {Koo}}{{Bershady}
  et~al.}{1998}]{bershady98}
{Bershady} M.~A.,  {Lowenthal} J.~D.,   {Koo} D.~C.,  1998, \mndoi [\apj]
  {10.1086/306130}, \href {http://adsabs.harvard.edu/abs/1998ApJ...505...50B}
  {505, 50}

\bibitem[\protect\citeauthoryear{{Bertin} \& {Arnouts}}{{Bertin} \&
  {Arnouts}}{1996}]{sextractor}
{Bertin} E.,  {Arnouts} S.,  1996, \mndoi [\aaps] {10.1051/aas:1996164}, \href
  {http://adsabs.harvard.edu/abs/1996A%26AS..117..393B} {117, 393}

\bibitem[\protect\citeauthoryear{{Bouwens} et~al.,}{{Bouwens}
  et~al.}{2014}]{bouwens2014}
{Bouwens} R.~J.,  et~al., 2014, \mndoi [\apj] {10.1088/0004-637X/793/2/115},
  \href {http://adsabs.harvard.edu/abs/2014ApJ...793..115B} {793, 115}

\bibitem[\protect\citeauthoryear{{Bouwens} et~al.,}{{Bouwens}
  et~al.}{2015}]{bouwens2015}
{Bouwens} R.~J.,  et~al., 2015, \mndoi [\apj] {10.1088/0004-637X/803/1/34},
  \href {http://adsabs.harvard.edu/abs/2015ApJ...803...34B} {803, 34}

\bibitem[\protect\citeauthoryear{{Bouwens} et~al.,}{{Bouwens}
  et~al.}{2016}]{bouwens2016}
{Bouwens} R.~J.,  et~al., 2016, \mndoi [\apj] {10.3847/0004-637X/830/2/67},
  \href {http://adsabs.harvard.edu/abs/2016ApJ...830...67B} {830, 67}

\bibitem[\protect\citeauthoryear{{Bowler} et~al.,}{{Bowler}
  et~al.}{2014}]{bowler2014}
{Bowler} R.~A.~A.,  et~al., 2014, \mndoi [\mnras] {10.1093/mnras/stu449}, \href
  {http://adsabs.harvard.edu/abs/2014MNRAS.440.2810B} {440, 2810}

\bibitem[\protect\citeauthoryear{{Bowler} et~al.,}{{Bowler}
  et~al.}{2015}]{bowler2015}
{Bowler} R.~A.~A.,  et~al., 2015, \mndoi [\mnras] {10.1093/mnras/stv1403},
  \href {http://adsabs.harvard.edu/abs/2015MNRAS.452.1817B} {452, 1817}

\bibitem[\protect\citeauthoryear{{Bradley} et~al.,}{{Bradley}
  et~al.}{2012}]{bradley2012}
{Bradley} L.~D.,  et~al., 2012, \mndoi [\apj] {10.1088/0004-637X/760/2/108},
  \href {http://adsabs.harvard.edu/abs/2012ApJ...760..108B} {760, 108}

\bibitem[\protect\citeauthoryear{Bradley et~al.,}{Bradley
  et~al.}{2016}]{photutils}
Bradley L.,  et~al., 2016, astropy/photutils: v0.3,
  \mndoi{10.5281/zenodo.164986}, \url {https://doi.org/10.5281/zenodo.164986}

\bibitem[\protect\citeauthoryear{{Calvi} et~al.,}{{Calvi}
  et~al.}{2016}]{calvi2016}
{Calvi} V.,  et~al., 2016, \mndoi [\apj] {10.3847/0004-637X/817/2/120}, \href
  {http://adsabs.harvard.edu/abs/2016ApJ...817..120C} {817, 120}

\bibitem[\protect\citeauthoryear{{Ciotti}}{{Ciotti}}{1991}]{ciotti91}
{Ciotti} L.,  1991, \aap, \href
  {http://adsabs.harvard.edu/abs/1991A%26A...249...99C} {249, 99}

\bibitem[\protect\citeauthoryear{{Coe}, {Ben{\'{\i}}tez}, {S{\'a}nchez}, {Jee},
  {Bouwens}  \& {Ford}}{{Coe} et~al.}{2006}]{coe2006}
{Coe} D.,  {Ben{\'{\i}}tez} N.,  {S{\'a}nchez} S.~F.,  {Jee} M.,  {Bouwens} R.,
    {Ford} H.,  2006, \mndoi [\aj] {10.1086/505530}, \href
  {http://adsabs.harvard.edu/abs/2006AJ....132..926C} {132, 926}

\bibitem[\protect\citeauthoryear{{Coe} et~al.,}{{Coe} et~al.}{2013}]{coe2013}
{Coe} D.,  et~al., 2013, \mndoi [\apj] {10.1088/0004-637X/762/1/32}, \href
  {http://adsabs.harvard.edu/abs/2013ApJ...762...32C} {762, 32}

\bibitem[\protect\citeauthoryear{{Crist{\'o}bal-Hornillos}
  et~al.,}{{Crist{\'o}bal-Hornillos} et~al.}{2009}]{hornillos2009}
{Crist{\'o}bal-Hornillos} D.,  et~al., 2009, \mndoi [\apj]
  {10.1088/0004-637X/696/2/1554}, \href
  {http://adsabs.harvard.edu/abs/2009ApJ...696.1554C} {696, 1554}

\bibitem[\protect\citeauthoryear{{Egami} et~al.,}{{Egami} et~al.}{2010}]{hls}
{Egami} E.,  et~al., 2010, \mndoi [\aap] {10.1051/0004-6361/201014696}, \href
  {http://adsabs.harvard.edu/abs/2010A%26A...518L..12E} {518, L12}

\bibitem[\protect\citeauthoryear{{Giavalisco} et~al.,}{{Giavalisco}
  et~al.}{2004}]{goods}
{Giavalisco} M.,  et~al., 2004, \mndoi [\apjl] {10.1086/379232}, \href
  {http://adsabs.harvard.edu/abs/2004ApJ...600L..93G} {600, L93}

\bibitem[\protect\citeauthoryear{{Graham} \& {Driver}}{{Graham} \&
  {Driver}}{2005}]{graham2005}
{Graham} A.~W.,  {Driver} S.~P.,  2005, \mndoi [\pasa] {10.1071/AS05001}, \href
  {http://adsabs.harvard.edu/abs/2005PASA...22..118G} {22, 118}

\bibitem[\protect\citeauthoryear{{Hathi} et~al.,}{{Hathi}
  et~al.}{2010}]{hathi2010}
{Hathi} N.~P.,  et~al., 2010, \mndoi [\apj] {10.1088/0004-637X/720/2/1708},
  \href {http://adsabs.harvard.edu/abs/2010ApJ...720.1708H} {720, 1708}

\bibitem[\protect\citeauthoryear{{H{\"a}ussler} et~al.,}{{H{\"a}ussler}
  et~al.}{2007}]{haussler2007}
{H{\"a}ussler} B.,  et~al., 2007, \mndoi [\apjs] {10.1086/518836}, \href
  {http://adsabs.harvard.edu/abs/2007ApJS..172..615H} {172, 615}

\bibitem[\protect\citeauthoryear{{H{\"a}u{\ss}ler} et~al.,}{{H{\"a}u{\ss}ler}
  et~al.}{2013}]{haussler2013}
{H{\"a}u{\ss}ler} B.,  et~al., 2013, \mndoi [\mnras] {10.1093/mnras/sts633},
  \href {http://adsabs.harvard.edu/abs/2013MNRAS.430..330H} {430, 330}

\bibitem[\protect\citeauthoryear{{Imai}, {Matsuhara}, {Oyabu}, {Wada},
  {Takagi}, {Fujishiro}, {Hanami}  \& {Pearson}}{{Imai}
  et~al.}{2007}]{imai2007}
{Imai} K.,  {Matsuhara} H.,  {Oyabu} S.,  {Wada} T.,  {Takagi} T.,  {Fujishiro}
  N.,  {Hanami} H.,   {Pearson} C.~P.,  2007, \mndoi [\aj] {10.1086/513513},
  \href {http://adsabs.harvard.edu/abs/2007AJ....133.2418I} {133, 2418}

\bibitem[\protect\citeauthoryear{{Ishigaki}, {Kawamata}, {Ouchi}, {Oguri}  \&
  {Shimasaku}}{{Ishigaki} et~al.}{2017}]{ishigaki2017}
{Ishigaki} M.,  {Kawamata} R.,  {Ouchi} M.,  {Oguri} M.,   {Shimasaku} K.,
  2017, preprint, \href {http://adsabs.harvard.edu/abs/2017arXiv170204867I} {}
  (\mn@eprint {arXiv} {1702.04867})

\bibitem[\protect\citeauthoryear{{Jiang} et~al.,}{{Jiang}
  et~al.}{2011}]{jian2011}
{Jiang} L.,  et~al., 2011, \mndoi [\apj] {10.1088/0004-637X/743/1/65}, \href
  {http://adsabs.harvard.edu/abs/2011ApJ...743...65J} {743, 65}

\bibitem[\protect\citeauthoryear{{Koekemoer}, {Fruchter}, {Hook}  \&
  {Hack}}{{Koekemoer} et~al.}{2003}]{multidrizzle}
{Koekemoer} A.~M.,  {Fruchter} A.~S.,  {Hook} R.~N.,   {Hack} W.,  2003, in
  {Arribas} S.,  {Koekemoer} A.,   {Whitmore} B.,  eds, HST Calibration
  Workshop : Hubble after the Installation of the ACS and the NICMOS Cooling
  System. p.~337

\bibitem[\protect\citeauthoryear{{Koekemoer} et~al.,}{{Koekemoer}
  et~al.}{2011}]{candels}
{Koekemoer} A.~M.,  et~al., 2011, \mndoi [\apjs] {10.1088/0067-0049/197/2/36},
  \href {http://adsabs.harvard.edu/abs/2011ApJS..197...36K} {197, 36}

\bibitem[\protect\citeauthoryear{{Lotz} et~al.,}{{Lotz} et~al.}{2017}]{hstff}
{Lotz} J.~M.,  et~al., 2017, \mndoi [\apj] {10.3847/1538-4357/837/1/97}, \href
  {http://adsabs.harvard.edu/abs/2017ApJ...837...97L} {837, 97}

\bibitem[\protect\citeauthoryear{{Mobasher} et~al.,}{{Mobasher}
  et~al.}{2005}]{mobasher2005}
{Mobasher} B.,  et~al., 2005, \mndoi [\apj] {10.1086/497626}, \href
  {http://adsabs.harvard.edu/abs/2005ApJ...635..832M} {635, 832}

\bibitem[\protect\citeauthoryear{{Oesch} et~al.,}{{Oesch}
  et~al.}{2014}]{oesch2014}
{Oesch} P.~A.,  et~al., 2014, \mndoi [\apj] {10.1088/0004-637X/786/2/108},
  \href {http://adsabs.harvard.edu/abs/2014ApJ...786..108O} {786, 108}

\bibitem[\protect\citeauthoryear{{Oesch} et~al.,}{{Oesch}
  et~al.}{2016}]{oesch2016}
{Oesch} P.~A.,  et~al., 2016, \mndoi [\apj] {10.3847/0004-637X/819/2/129},
  \href {http://adsabs.harvard.edu/abs/2016ApJ...819..129O} {819, 129}

\bibitem[\protect\citeauthoryear{{Peng}, {Ho}, {Impey}  \& {Rix}}{{Peng}
  et~al.}{2002}]{peng2002}
{Peng} C.~Y.,  {Ho} L.~C.,  {Impey} C.~D.,   {Rix} H.-W.,  2002, \mndoi [\aj]
  {10.1086/340952}, \href {http://adsabs.harvard.edu/abs/2002AJ....124..266P}
  {124, 266}

\bibitem[\protect\citeauthoryear{{Postman} et~al.,}{{Postman}
  et~al.}{2012}]{clash}
{Postman} M.,  et~al., 2012, \mndoi [\apjs] {10.1088/0067-0049/199/2/25}, \href
  {http://adsabs.harvard.edu/abs/2012ApJS..199...25P} {199, 25}

\bibitem[\protect\citeauthoryear{{Schmidt} et~al.,}{{Schmidt}
  et~al.}{2014}]{schmidt2014}
{Schmidt} K.~B.,  et~al., 2014, \mndoi [\apj] {10.1088/0004-637X/786/1/57},
  \href {http://adsabs.harvard.edu/abs/2014ApJ...786...57S} {786, 57}

\bibitem[\protect\citeauthoryear{{S\'ersic}}{{S\'ersic}}{1968}]{sersic68}
{S\'ersic} J.~L.,  1968, {Atlas de galaxias australes}

\bibitem[\protect\citeauthoryear{{Stanway}, {Bremer}  \& {Lehnert}}{{Stanway}
  et~al.}{2008}]{stanway2008}
{Stanway} E.~R.,  {Bremer} M.~N.,   {Lehnert} M.~D.,  2008, \mndoi [\mnras]
  {10.1111/j.1365-2966.2008.12853.x}, \href
  {http://adsabs.harvard.edu/abs/2008MNRAS.385..493S} {385, 493}

\bibitem[\protect\citeauthoryear{{Steidel}, {Giavalisco}, {Dickinson}  \&
  {Adelberger}}{{Steidel} et~al.}{1996}]{steidel96}
{Steidel} C.~C.,  {Giavalisco} M.,  {Dickinson} M.,   {Adelberger} K.~L.,
  1996, \mndoi [\aj] {10.1086/118019}, \href
  {http://adsabs.harvard.edu/abs/1996AJ....112..352S} {112, 352}

\bibitem[\protect\citeauthoryear{{Trenti}, {Stiavelli}, {Bouwens}, {Oesch},
  {Shull}, {Illingworth}, {Bradley}  \& {Carollo}}{{Trenti}
  et~al.}{2010}]{trenti2010}
{Trenti} M.,  {Stiavelli} M.,  {Bouwens} R.~J.,  {Oesch} P.,  {Shull} J.~M.,
  {Illingworth} G.~D.,  {Bradley} L.~D.,   {Carollo} C.~M.,  2010, \mndoi
  [\apjl] {10.1088/2041-8205/714/2/L202}, \href
  {http://adsabs.harvard.edu/abs/2010ApJ...714L.202T} {714, L202}

\bibitem[\protect\citeauthoryear{{Trenti} et~al.,}{{Trenti}
  et~al.}{2011}]{trenti2011}
{Trenti} M.,  et~al., 2011, \mndoi [\apjl] {10.1088/2041-8205/727/2/L39}, \href
  {http://adsabs.harvard.edu/abs/2011ApJ...727L..39T} {727, L39}

\bibitem[\protect\citeauthoryear{{Trenti} et~al.,}{{Trenti}
  et~al.}{2012}]{trenti2012}
{Trenti} M.,  et~al., 2012, \mndoi [\apj] {10.1088/0004-637X/746/1/55}, \href
  {http://adsabs.harvard.edu/abs/2012ApJ...746...55T} {746, 55}

\bibitem[\protect\citeauthoryear{{Williams} et~al.,}{{Williams}
  et~al.}{1996a}]{hdf}
{Williams} R.~E.,  et~al., 1996a, \mndoi [\aj] {10.1086/118105}, \href
  {http://adsabs.harvard.edu/abs/1996AJ....112.1335W} {112, 1335}

\bibitem[\protect\citeauthoryear{{Williams} et~al.,}{{Williams}
  et~al.}{1996b}]{williams96}
{Williams} R.~E.,  et~al., 1996b, \mndoi [\aj] {10.1086/118105}, \href
  {http://adsabs.harvard.edu/abs/1996AJ....112.1335W} {112, 1335}

\bibitem[\protect\citeauthoryear{{Yan} et~al.,}{{Yan} et~al.}{2011}]{yan2011}
{Yan} H.,  et~al., 2011, \mndoi [\apjl] {10.1088/2041-8205/728/1/L22}, \href
  {http://adsabs.harvard.edu/abs/2011ApJ...728L..22Y} {728, L22}

\bibitem[\protect\citeauthoryear{{Zitrin} et~al.,}{{Zitrin}
  et~al.}{2014}]{zitrin2014}
{Zitrin} A.,  et~al., 2014, \mndoi [\apjl] {10.1088/2041-8205/793/1/L12}, \href
  {http://adsabs.harvard.edu/abs/2014ApJ...793L..12Z} {793, L12}

\bibitem[\protect\citeauthoryear{{de Vaucouleurs}}{{de
  Vaucouleurs}}{1948}]{devaucouleurs}
{de Vaucouleurs} G.,  1948, Annales d'Astrophysique, \href
  {http://adsabs.harvard.edu/abs/1948AnAp...11..247D} {11, 247}

\makeatother
\end{thebibliography}

\end{document}